\documentstyle[fleqn,11pt]{article}
\begin{document}
\bigskip
\parindent 1.4cm
\large

\begin{center}
{\Large \bf Wave Packet under Continuous Measurement 
\par via Bohmian Mechanics}
\end{center}
\vspace{1.0cm}
\begin{center}
{Ant\^onio B. Nassar} 
\end{center}
\begin{center}
{\it Physics Department
\par The Harvard-Westlake School
\par 3700 Coldwater Canyon, Studio City, 91604 (USA)
\par and
\par Department of Sciences 
\par University of California, Los Angeles, Extension Program
\par 10995 Le Conte Avenue, Los Angeles, CA 90024 (USA)
}
\end{center}
\vspace{1.0cm}
\par

\begin{center}
{\bf Abstract}
\end{center}

A new quantum mechanical description of the dynamics of wave packet under continuous measurement is formulated via Bohmian mechanics. The solution to this equation is found through a wave packet approach which establishes a direct correlation between a classical variable with a quantum variable describing the dynamics of the center of mass and the width of the wave packet. The approach presented in this paper gives a comparatively clearer picture than approaches using restrited path integrals and master equation approaches. This work shows how the extremely irregular character of classical chaos can be reconciled with the smooth and wavelike nature of phenomena on the atomic scale. It is demonstrated that a wave packet under continuous quantum measurement displays both chaotic and non-chaotic features. The Lyapunov characteristic exponents for the trajectories of classical particle and the quantum wave packet center of mass are calculated and their chaoticities are demonstrated to be about the same. Nonetheless, the width of the wave packet exhibits a non-chaotic behavior and allows for the possibility to beat the standard quantum limit by means of transient, contractive states.

\vspace{1.5cm}
\vspace{1.5cm}
\pagebreak

The evolution and dynamics of wave packets are the subject of much current investigation in many areas of both physics and chemistry.\cite{yeazell,nauenberg} In molecular physics, a new realm of phenomena involving wave packets has opened up with the emergence of femtosecond pulse technology.\cite{garraway} Wave packets have also been produced in semiconductor quantum well systems.\cite{leo} The use of wave packets to analyze the dynamics of quantum mechanical systems is crucial for the study of the classical-quantum interface. Further, the description of a real scattering event should correspond to the established experimental observation of localized particles approaching a scattering center and subsequently receding from it. This entails the construction of a wave packet state and the analysis of its evolution in time.

A new quantum mechanical description of the dynamics of wave packet under continuous measurement is formulated via Bohmian mechanics. The solution to this equation is found through a wave packet approach which establishes a direct correlation between a classical variable with a quantum variable describing the dynamics of the center of mass and the width of the wave packet. The approach presented in this paper gives a comparatively clearer picture than approaches using restrited path integrals and master equation approaches. This work shows how the extremely irregular character of classical chaos can be reconciled with the smooth and wavelike nature of phenomena on the atomic scale. It is demonstrated that a wave packet under continuous quantum measurement displays both chaotic and non-chaotic features. The Lyapunov characteristic exponents for the trajectories of classical particle and the quantum wave packet center of mass are calculated and their chaoticities are demonstrated to be about the same. Nonetheless, the width of the wave packet exhibits a non-chaotic behavior and allows for the possibility to beat the standard quantum limit by means of transient, contractive states.\cite{yuen} In particular, the periodically driven Duffing oscillator, which has become a classic model for analysis of nonlinear phenomena,\cite{habib,brun,strunz} is studied, and its classical chaos is shown to crossover into the quantum regime. The theory presented below focuses on some unresolved features posed by chaos and on the correspondence principle which is the main focus of many new experiments with excited atomic and molecular systems. These experiments can directly probe the realm of high quantum numbers or classically chaotic motion. 

The renewed interest in coupling classical systems to quantum ones has been revived by a number of authors\cite{bhattacharya,diosi,mensky} who have examined continuous quantum measurements. The question of coupling classical variables to quantum variables is intimately connected to the question of how certain variables become classical in the first place.\cite{caron et al1} In reality, there are no fundamentally classical systems, only quantum systems that are effectively classical under certain conditions. One must start from the underlying quantum theory of the whole composite system and then derive the effective form of the classical theory. The starting point is to think of the classical particle as continuously monitoring the quantum particle's position and responding to the measured value. To this end, consider a classical particle of mass $M$ with position $X$ in a nonlinear potential, the periodically driven Duffing oscillator, coupled to a quantum oscillator of frequency $\omega$ and mass $m$:
\begin{eqnarray} \label{eq:classicalX}
M\ddot X(t) + BX^3 (t) - AX(t) + \lambda \bar x(t) = \Lambda \cos \left( {\Omega t} \right),
\end{eqnarray}
where $\bar x(t)$ is associated with the measurement record of the quantum system. The evolution of the wave function of the quantum system $\psi$ can be expressed at first in terms of the path-integral for the unnormalized wave function:
\begin{eqnarray} \label{eq:pathintegral}
\psi (x',t') = \int {\cal D} [x(t)] \exp \left[ {{i \over \hbar }\int\limits_0^{t'} {dt\left( {{1 \over 2}m\dot x^2(t)  - {1 \over 2}m\omega ^2 x^2(t)  - \lambda x(t)X(t)} \right)} } \right] \nonumber \\ \times 
\exp \left( {- \int\limits_0^{t'} {dt} {{\left[x(t) - \bar x(t)\right]^2 } \over {4\sigma ^2(t) }}} \right)\psi (x_0 ,0),
\end{eqnarray}
where the path integral is over paths $x(t)$ satisfying $x(0) = x_0$ and $x(t') = x'$. The quantity $\sigma$ in the equation above represents the resolution of the effective measurement of the particle by the classical system, as indicated by previous works.\cite{bhattacharya,diosi,mensky} However, some differences here are worth mentioning. One is the time dependence of the quantity $\sigma(t)$: most importantly is the novelty that the general resolution of the measurement evolves according to a nonlinear differential equation. Another difference relates to the dimension of the quantity $\sigma(t)$: it should be considered only proportional to the actual position uncertainty in the measurement of the quantum particle. So, an explicit connection to a wave packet approach can be established by writing $\sigma^2(t) = \tau \delta^2(t)$, where $\delta$ and $\tau$ have dimensions of space and time, respectively. This point can be further elucidated by approximating the last term of Equation (\ref{eq:pathintegral}) around an average time $\bar t$, i. e., $ \sim \exp  - [(x(\bar t) - \bar x(\bar t))^2 /4\delta (\bar t)^2 ]\exp ( - \bar t/\tau )$, where $\delta(t)$ clearly stands for the position uncertainty (width of the wave packet) and $\tau$ characterizes the time constant (relaxation time) of the measurement. 

Now, the square of the absolute value of Equation (\ref{eq:pathintegral}) yields the probability density for different measurement outputs at different times and from this equation the associated Schr\"odinger equation describing the system undergoing continuous measurement can be written as:
\begin{eqnarray} \label{eq:schroedinger}
i\hbar {{\partial \psi (x,t)} \over {\partial t}} =  - {{\hbar ^2 } \over {2m}}{{\partial ^2 \psi (x,t)} \over {\partial x^2 }} + \left( {{1 \over 2}m\omega ^2 x^2  + \lambda xX(t)} \right)\psi (x,t) \nonumber \\
- {{i\hbar } \over {4 \tau}}\left( {{{\left[ {x - \bar x(t)} \right]^2 } \over {\delta ^2 (t)}} - 1} \right)\psi (x,t).
\end{eqnarray}

Next, a solution to this equation can be found by considering previous findings \cite{strunz,bhattacharya} which have shown that continuous position measurement produces and maintains localization in phase space as a necessary result of the information it provides. In addition to localizing the state, a continuous position measurement can also introduce noise in its evolution: the measured value $\bar x(t)$ can be associated with a mean value $<x(t)>$ plus a noise-dependent component $\xi(t)$. So, in order to obtain a semiclassical protocol one must be in a regime in which the localization is relatively strong and the noise sufficiently weak. However, a protocol based on a complete hierarchy of stochastic equations associated with the average value of the position $<x(t)>$ makes it difficult to obtain an analytic solution to the problem.\cite{bhattacharya,note1} The details of the variances and resulting noise strength permit only partial solutions based on varying $\hbar$ and steady state regimes. Therefore, a formalism that keeps the measurement record quantity $\bar x(t)$ without dealing with the details of the variances can circumvent this difficult task and give a direct description of the evolution of the quantum system. This rationale entails a wave packet solution around the measurement record $\bar x(t)$ as follows:
\begin{eqnarray} \label{eq:wavepacket}
\left| {\psi (x,t)} \right| = \left[ {2\pi \delta^2 (t)} \right]^{ - 1/4} \exp \left( -{ {{[x - \bar x(t)]^2 } \over {4\delta^2 (t)}}} \right).
\end{eqnarray}

This minimum-uncertainty wave packet solution is further supported by recent, alternative stochastic approaches\cite{strunz} which have demonstrated that individual quantum trajectories remain minimum-uncertainty localized wave packets for all times: the localization being stronger the smaller $\hbar$ becomes. Similar localization properties hold also for a variety of quantum trajectory methods\cite{gisin,schack,percival} where the mean uncertainty product $M[\Delta x \Delta p]/\hbar$ remains close to 1 almost independent of $\hbar$, thus corroborating the minimum-uncertainty ansatz ($\ref{eq:wavepacket}$). These quantum trajectory methods have been used extensively in recent years due their intimate connection to continuous measurement.

Within the Bohmian mechanics\cite{tumulka}-\cite{bell} a framework for analyzing quantum trajectories is provided by assuming that the wave function which satisfies Schr\"odinger's equation is no longer the most complete description of the state of the system. It ascribes a particle motion via the de Broglie guidance condition
\begin{eqnarray} \label{eq:debroglie}
v(x,t) = \frac{1}{m}\frac{{\partial S(x,t)}}{{\partial x}}
\end{eqnarray}
where $v$ represents the particle velocity and $S$ is the phase of the wave function $\psi$. By expressing the wave function in polar form as 
\begin{eqnarray} \label{eq:wavefunction}
\psi(x,t)  = \phi(x,t) \exp (iS(x,t)/\hbar ),
\end{eqnarray}
Schr\"odinger's equation can be recast as
\begin{eqnarray} \label{eq:partialv}
{{\partial v} \over {\partial t}} + v{{\partial v} \over {\partial x}} =  - {1 \over m}{\partial  \over {\partial x}}\left( {V_{ext}  + V_{qu} } \right),
\end{eqnarray}
and
\begin{eqnarray} \label{eq:partialrho}
{{\partial \rho } \over {\partial t}} + {\partial  \over {\partial x}}\left( {\rho v} \right) = 0.
\end{eqnarray}.

Equation (\ref{eq:partialv}) can be regarded as a modified Hamilton-Jacobi equation while Equation (\ref{eq:partialrho}) is a continuity equation for $\rho  = \phi ^2$; $V_{ext}$ denotes the external classical potential and 
\begin{eqnarray} \label{eq:quV}
V_{qu}  =  - {\rm{ }}{{\hbar ^2 } \over {2m{\rm{ }}\phi }}{{\partial ^2 \phi } \over {\partial x^2 }}
\end{eqnarray}
is the so-called quantum potential.

Now, by substituting of Equations (\ref{eq:wavefunction}) and (\ref{eq:wavepacket}) into Equation (\ref{eq:schroedinger}),\cite{nassar} the auxiliary functions of time $\delta(t)$ and $\bar x(t)$ of the wave packet can be shown to conform to the following equations:
\begin{eqnarray} \label{eq:quantumxeqn}
\ddot {\bar x}(t) + \omega ^2 \bar x(t) + \left( {{\lambda  \over m}} \right)X(t) = 0.
\end{eqnarray}
and
\begin{eqnarray} \label{eq:deltaeqn}
\ddot \delta(t)  + {1 \over \tau}\dot \delta(t)  + \left( {\omega ^2  + {1 \over {4\tau^2 }}} \right)\delta(t)  = {{\hbar ^2 } \over {4m^2 \delta ^3(t) }}
\end{eqnarray}

Equations (\ref{eq:quantumxeqn}) and (\ref{eq:deltaeqn}) show that a continuous measurement of a quantum oscillator gives specific features to its evolution: the appearance of distinct classical and quantum elements. This measurement consists of monitoring the position of the quantum system and the result is the measured path $\bar x(t)$ for $t$ within an uncertainty $\delta(t)$. The solutions to these equations are presented as follows. 

First, Equation (\ref{eq:quantumxeqn}) demonstrates the claim that continuous measurement can effectively obtain classical mechanics from quantum mechanics. The Lyapunov exponents that separate different time scales of motion are established for both classical and quantum solutions as follows:
\begin{eqnarray} \label{eq:lambda}
\lambda _{(cl,qu)} = \mathop {\lim }\limits_{\scriptstyle t \to \infty  \hfill \atop 
  \scriptstyle \Delta (0) \to 0 \hfill}  \left\{ {\ln \left[ {\Delta _{(cl , qu) } (t)/\Delta _{(cl , qu)} (0)} \right]/t} \right\},
\end{eqnarray}
where 
\begin{eqnarray} \label{eq:delta}
\Delta _{(cl,qu)} (t) = \left\{ {[(X^ +, \bar x ^ +)  - (X^ -, \bar x ^ -) ]^2  + [(\dot X ^ +, \dot {\bar x} ^ +)   - (\dot X ^ -, \dot {\bar x} ^ - )   ]^2 } \right\}^{1/2}
\end{eqnarray}
represents the renormalized classical, quantum Euclidean distances of the trajectories in phase space, respectively. Equation (\ref{eq:lambda}) describes explicitly the asymptotic rate of exponential divergence of the classical and quantum trajectories evolving from two initially close initial conditions, respectively. It appears that, at least initially, the logarithmic divergence of trajectories with a very small perturbation in the initial conditions is roughly linear on this plot, indicating an exponential relationship. To find the exponent we need to find a line that fits the logarithm of the data. Thus, it is appropriate to use only the data up to the point where the difference is of order one. Although a perturbation causes exponential divergence locally, solutions near this initial condition are attracted to a strange attractor, which is a bounded set with zero area. Since this set is bounded, the divergence cannot continue indefinitely. A regression on the data gets us a reasonable exponential function to model the divergence: for the classical case, $8 \times 10^{ - 7} e^{0.17(1)t} $ and for the quantum case $5 \times 10^{ - 7} e^{0.16(8)t} $. Thus, the behavior of a quantum wave packet center of mass and the monitoring classical coordinate are equally chaotic and the Lyapunov exponents for both cases is found to be: $\lambda_{qu} \simeq \lambda_{cl} = 0.17$.

On the other hand, Equation (\ref{eq:deltaeqn}) shows that the width of the wave packet exhibits a non-chaotic behavior. In this context, a solution to Equation (\ref{eq:deltaeqn}) for a free particle ($\omega = 0$) supports qualitatively Yuen's conclusions \cite{yuen} so far as showing the possibility to beat the standard quantum limit by means of transient, contractive states. Extensive deliberations on how to defend or beat the standard quantum limit for both discrete and continuous measurements of the position of a quantum particle can be found in the literature.\cite{mabuchi}-\cite{braginski} Accurate measurements of the position of a particle is of much interest in the context of gravitational-wave detection where questions have arisen as to whether there are fundamental quantum mechanical limits on detection sensitivity. The point here is that discrete or continuous measurements may introduce squeezing that affects subsequent measurements. Besides, the resolution squared [$\sigma^2(t) = \tau \delta^2(t)$] of the measurement can reach a stationary regime, namely:
\begin{eqnarray} \label{eq:sigmadelta}
\sigma _o^2  = {{\hbar \tau ^2 /m} \over {\left( {1 + 4\omega ^2 \tau ^2 } \right)^{1/2} }},
\end{eqnarray}
which indicates that localization can occur on a time scale which might be extremely short compared to the oscillator's frequency $\omega$. For the low-frequency limit $\omega \tau  \ll 1$ (the free particle limit $\omega = 0$), this result reduces to $\sigma _o^2  = \hbar \tau^2 /m $. On the other hand, for the high-frequency limit $\omega \tau  \gg 1$, $\sigma _o^2 = \hbar \tau /2m\omega $. These results show that the resolution $\sigma$ of the effective measurement increases as the characteristic time constant $\tau$ (relaxation time) increases.

To conclude, this work has developed a wave packet approach from a path-integral formalism to describe a continuously measured quantum particle's position by a classical particle and to establish a direct correlation between a classical variable $X$ with a quantum variable $\bar x$. 
It shows how the extremely irregular character of classical chaos can be reconciled with the smooth and wavelike nature of phenomena by demonstrating that a wave packet under continuous quantum measurement displays both chaotic and non-chaotic features. The Lyapunov characteristic exponents for the trajectories of classical particle and the quantum wave packet center of mass are calculated and their chaoticities are demonstrated to be about the same. On the other hand, the width of the wave packet exhibits a non-chaotic behavior and allows for the possibility to beat the standard quantum limit by means of transient, contractive states.

{\bf Acknowledgments}

It is a pleasure to acknowledged comments on this work by J. Patterson and M. B. Mensky.


\newpage

\begin{thebibliography}{99}

\bibitem{yeazell} J. A. Yeazell and T. Uzer, {\it The Physics and Chemistry of Wave Packets} (Wiley, N. Y., 2000).

\bibitem{nauenberg} M. Nauenberg, C. Stroud, and J. Yeazell, {\it Sci. Am.} {\bf 270} (1994) 44.

\bibitem{garraway} B. M. Garraway and K-A. Suominen, {\it Rep. Prog. Phys.} {\bf 58} (1995) 365.

\bibitem{leo} K. Leo, J. Shah, E. O. Gobel, and T. C. Damen, {\it Phys. Rev. Lett.}, {\bf 66} (1991) 201.

\bibitem{yuen} H. P. Yuen, {\it Phys. Rev. Lett.} {\bf 51} (1983) 719;{\it ibid.} {\bf 52} (1984) 1730.

\bibitem{habib} S. Habib, K. Jacobs, H. Mabuchi, R. Ryne, K. Shizume, and B. Sundaram, {\it Phys. Rev. Lett.} {\bf 88} (2002) 1.

\bibitem{brun} T. A. Brun, N. Gisin, P. F. O'Mahony, and M. Rigo, {\it Phys. Lett.} {\bf A229} (1997) 267.

\bibitem{strunz} W. T. Strunz, L. Di\'osi, N. Gisin, and T Yu, {\it Phys. Rev. Lett.} {\bf 83} (1999) 4909 and references therein.

\bibitem{bhattacharya} T. Bhattacharya, S. Habib, and K. Jacobs, {\it Phys. Rev. Lett.} {\bf 85} (2000) 4852. See also references therein.

\bibitem{diosi} L. Di\'osi and J. J. Halliwell, {\it Phys. Rev. Lett.} {\bf 81} (1998) 2846.

\bibitem{mensky} M. B. Mensky, {\it Phys. Lett.} {\bf A307} (2003) 85;{\it Phys. Lett.} {\bf A231} (1997) 1; {\it Int. J. Theor. Phys.} {\bf 37} (1998) 273;{\it Continuous Quantum Measurement and Path Integrals} (IOP Pub., Bristol, 1993).

\bibitem{caron et al1} L. A. Caron, D. Huard, H. Kr\"oger, G. Melkonyan, K. J. M. Moriarty and L. P. Nadeau, {\it Phys. Lett.} {\bf A322} (2004) 60.

\bibitem{note1} If the noise is significant, thermal effects compete with quantum ones. To obtain the average probability solution for this process, use can be made  of  Chandrasekhar's convolution lemma: $ < p(x,t) >  = \int_{ - \infty }^{ + \infty } {p(x -  < x > ,t)W( < x > ,t)d < x > } $, where $p (x,t) = \left| {\psi (x,t)} \right| ^2$ and $W(<x>,t)$ is the thermal probability of some value of $<x>$ at time $t$ for given initial conditions. See S. Chandrasekhar, {\it Rev. Mod. Phys.} {\bf 15} (1943) 43 and A. B. Nassar, {\it Phys. Rev. A} {\bf 33} (1986) 2134. A detailed discussion of this procedure will be given in a more extended publication.

\bibitem{gisin} N. Gisin and I.C. Percival, {\it J. Phys.} {\bf A26} (1993) 2233.

\bibitem{schack} R. Schack, T. Brun and I. C. Percival, {\it J. Phys.} {\bf A28} (1995) 5401.

\bibitem{percival} W. T. Strunz and I. C. Percival, {\it J. Phys.} {\bf A31} (1998) 1801. 

\bibitem{tumulka} R. Tumulka, {\it Am. J. Phys.} 72, 1220 (2004). See references therein.
\bibitem{oleg} Oleg V. Prezhdo and Craig Brooksby, {\it Phys. Rev. Lett.} 86, 3215 (2001).
\bibitem{nogamia} Y. Nogamia, F. M. Toyamab and W. van Dijk, {\it Phys. Lett.} A270, 279 (2000).
\bibitem{plastino} A. R. Plastino, M. Casas and A. Plastino {\it Phys. Lett.} A281, 297(2001). 
\bibitem{dias} N. C. Dias  and J. N. Prata {\it Phys. Lett.} A302, 261 (2002).
\bibitem{makowski} A. J. Makowski, , P. Pepowski and S. T. Dembiski {\it Phys. Lett.} A266, 241 (2000).
\bibitem{shifren} L. Shifren, , R. Akis and D. K. Ferry {\it Phys. Lett.} A274, 75 (2000).
\bibitem{falsaperla} P. Falsaperla and G. Fonte {\it Phys. Lett.} A316, 382 (2003).
\bibitem{datta} A. Datta, P. Ghose and M. K. Samal {\it Phys. Lett.} A322, 277 (2004).
\bibitem{colijn} C. Colijn and E. R. Vrscay {\it Phys. Lett.} A300, 334 (2002).
\bibitem{potel} G. Potel, M. Muñoz-Aleñar, F. Barranco and E. Vigezzi {\it Phys. Lett.} A299, 125 (2002).
\bibitem{stomphorst} R. G. Stomphorst {\it Phys. Lett.} A292, 213 (2002).
\bibitem{ali} Md. M. Ali, A. S. Majumdar and D. Home {\it Phys. Lett.} A304, 61 (2002).
\bibitem{bowman} G. E. Bowman {\it Phys. Lett.} A298, 7 (2002).
\bibitem{durr} D. D\"urr, S. Goldstein, R. Tumulka, and N. Zanghi, {\it Phys. Rev. Lett.} 
93, 090402 (2004). See references therein.
\bibitem{bell} J. S. Bell, {\it Speakable and Unspeakable in Quantum Mechanics} (Cambridge
University Press, Cambridge, 1987).

\bibitem{nassar} A precursor of this method can be found in: A. B. Nassar, {\it J. Opt. B: Quantum Semiclass. Opt.} {\bf 4} (2002) S226. 

\bibitem{mathematica} {\it Mathematica 6.0} (Wolfram Research, Inc., Champaign, IL, 2003).

\bibitem{mabuchi} H. Mabuchi, {\it Phys. Rev.} {\bf A58} (1998) 123.

\bibitem{ozawa} M. Ozawa, {\it Phys. Rev. Lett.} {\bf 60} (1988) 385.

\bibitem{ni} W. T. Ni, {\it Phys. Rev.} {\bf A33} (1986) 2225.

\bibitem{caves1} C. M. Caves, {\it Phys. Rev. Lett.} {\bf 54} (1985) 2465.

\bibitem{lynch} R. Lynch, {\it Phys. Rev. Lett.} {\bf 54} (1985) 1599.

\bibitem{braginski}V. B. Braginsky and F. Ya. Khalili, {\it Quantum Measumerent} (Cambridge Univ. Press, 1992); V. B. Braginsky and A. B. Manukin, {\it Measurement of Weak Forces in Physics Experiments} (Nauka, Moscow, 1974) [English translation edited
by D. H. Douglass (University of Chicago, Chicago, 1977)].

\end{thebibliography}
\end{document}